\documentclass[epj]{svjour}

\usepackage{graphicx}
\usepackage{fancyhdr}
\def\makeheadbox{{
 }}
\def\mytitle{My title} 
\def\myauthors{My name}  
\def\mytype{My type of session}
\def\mysession{My session}

\def\mytitle{Little Higgs, Non-standard Higgs, No Higgs and All That} 
\def\myauthors{Hsin-Chia Cheng}    
\def\mytype{Review}
\def\mysession{\myauthors}

\begin{document}
\title{Little Higgs, Non-standard Higgs, No Higgs and All That}
\author{Hsin-Chia Cheng\inst{1}
\thanks{\emph{Email:} cheng@physics.ucdavis.edu}%
}                     
%
%
\institute{Department of Physics, University of California, Davis, CA 95616, USA}
%
\date{}
\abstract{
We give a brief review of recent developments in non-supersymmetric models for electroweak symmetry breaking, including little Higgs, composite Higgs and Higgsless theories. The new ideas such as extra dimensions, AdS/CFT correspondence, dimension-deconstruction, and collective symmetry breaking provide us new tools to construct new models. They also allow some old ideas to be revived and implemented in these new models.
} 
\maketitle
\section{Introduction}
\label{intro}
The mechanism of electroweak symmetry breaking is currently one of the most prominent question in particle physics. In Standard Model (SM), electroweak symmetry is broken by a nonzero vacuum expectation value (vev) of a scalar Higgs field. There is an associated particle, the Higgs boson, which is responsible for unitarizing the longitudinal $WW$ scattering in the Higgs mechanism. The Higgs boson is the only missing piece in the Standard Model and the most hunted particle experimentally. However, a scalar field in general receives large quadratically divergent contributions to its mass-squared from its interactions and hence suffers from the hierarchy or naturalness problem: why is the Higgs mass and the scale of electroweak breaking so small compared with other fundamental scales such as the Planck scale or the grand unification scale? Because of the hierarchy problem, it is generally believed that there will be additional new physics at the electroweak scale which accompanies or replaces the Higgs field. 
The possible new physics will be tested at the Large Hadron Collider (LHC) which starts running in 2008. With a center of mass energy of 14 TeV, LHC is expected to have the ability to explore the TeV scale physics and address the question of electroweak symmetry breaking. However, the processes and the detections of new physics at the LHC experiments are very complex. We need to be ready for any possibility and know what we are looking for in order for the LHC to reach its ultimate discovering potential.

Supersymmetry (SUSY) has been the leading candidate for physics beyond the Standard Model and its experimental signatures have been extensively studied. However, there are also other well-motivated scenarios for TeV scale physics besides SUSY. In fact, there have been a lot of progresses in alternative theories in recent years based on new ideas such as (flat or warped) extra dimensions, collective symmetry breaking and AdS/CFT correspondence~\cite{Maldacena:1997re,Gubser:1998bc,Witten:1998qj,Arkani-Hamed:2000ds,Rattazzi:2000hs}. They allow us to construct new models or revive old ideas, calculate or estimate theoretical predictions, and find new ways to satisfy experimental constraints. The purpose of this talk is to give a brief review of the alternative scenarios to SUSY, with emphasis on new developments in recent years.

There are many challenges to face for any alternative theory at the TeV scale. As many alternative theories are based on strong dynamics, there are questions on theoretical consistency and predictivity:  how can we have theoretical control of the strong dynamics and how can we make predictions with any confidence? On the experimental side, though the reach of the direct search for new physics at the TeV scale is still quite limited, there are strong electroweak precision constraints from LEP, SLC, Tevatron, and other low energy experiments. Currently there is no significant deviation observed in comparison with the SM predictions. An important criterion in building new models at the TeV scale is to avoid large corrections to the electroweak observables in order to satisfy the experimental constraints.

There are too many models on the market and it is impossible to cover all of them. Many very different looking models in fact share similar ideas and can have similar phenomenology. I will divide these alternative models into the following broad categories:
\begin{itemize}
\item (Naturally) light Higgs: A naturally light Higgs can arise because its mass is protected by a shift symmetry. Examples are Higgs as a pseudo-Nambu-Goldstone boson (PNGB) or the extra component of a gauge field ($A_5$) in theories with extra dimensions. 

\item Heavy Higgs: Higgs arises as a composite field, but there is no extra symmetry to protect its mass. In this category, Higgs is expected to be heavy and close to the compositeness scale without fine tuning.

\item No Higgs: There is no physical Higgs boson. The strong $W_L\, W_L$ scattering is unitarized by some other states, e.g., techni-rhos in technicolor models and Kaluza-Klein (KK) gauge bosons in Higgsless Models.
\end{itemize}

\begin{figure}[htbp] 
   \centering
      \includegraphics[width=3in]{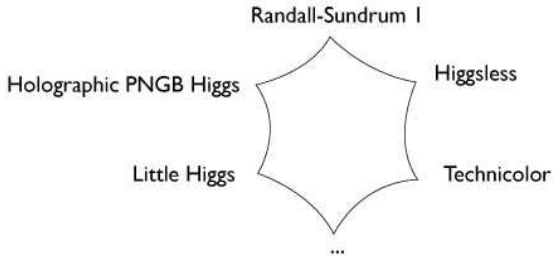} 
   \caption{The model space of alternative theories}
   \label{fig:m-theory}
\end{figure}

Such a division is somewhat artificial because the model space is really continuous (Fig.~\ref{fig:m-theory}). There are models which interpolate among these categories. Different models often share some similar features and face similar challenges. It is therefore desirable to have some uniform way to describe these models and their experimental signatures. To fully distinguish various models may require experimental information at very high energies, beyond the reach of the LHC.

\section{Light Higgs Scenario}
\label{sec:light_higgs}

The idea that a light Higgs can be a PNGB or $A_5$ has been around for a long time~\cite{Kaplan:1983fs,Kaplan:1983sm,Fairlie:1979mc,Manton:1979kb,Hosotani:1983xw}. Recent new ideas like dimension-deconstruction and collective symmetry breaking have allowed us to build more realistic models. They include
\begin{itemize}
\item Little Higgs models,
\item Gauge-Higgs unification based on flat or warped extra dimensions,
\item Twin Higgs models.
\end{itemize}
The twin Higgs models involve a hidden or mirror sector which only couples to the Standard Model through the Higgs. Their phenomenology can be quite challenging. They are very different from the other models and it's difficult to describe them together with the other models in a uniform way. In the following I will focus my talk on the other models. However, the twin Higgs models are an interesting class of models and I refer the interested readers to the original literature~\cite{Chacko:2005pe,Chacko:2005vw,Chacko:2005un}.

\subsection{Little Higgs Theories}

\begin{figure}[htbp] 
   \centering
      \includegraphics[width=3in]{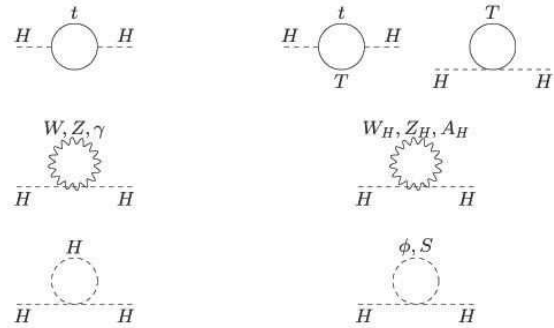} 
   \caption{The cancellation of one-loop quadratic divergences. the SM interactions are shown on the left side and the new particle interactions are on the right side.}
   \label{fig:cancellation}
\end{figure}

Little Higgs theories are a new type of theories which can stabilize the electroweak scale naturally~\cite{Arkani-Hamed:2001nc,Arkani-Hamed:2002qx,Arkani-Hamed:2002qy}. In little Higgs theories, Higgs fiels is a PNGB of some spontaneously broken global symmetry ($G\to H$). In addition, it has the special property of collective symmetry breaking. The global symmetry is explicitly broken by 2 sets of interactions, with each set preserving a subset of the symmetry,
\begin{equation}
{\cal L}= {\cal L}_0 + \lambda_1 {\cal L}_1 + \lambda_2 {\cal L}_2,
\end{equation}
where ${\cal L}_0$ is the symmetric part of the Lagrangian and ${\cal L}_1$ and ${\cal L}_2$ are explicit symmetry-breaking terms.
Higgs field is an exact Nambu-Goldstone boson when either set of couplings vanish, so a Higgs mass can only be generated in the presence of both sets of couplings. In this way, Higgs mass-squared are suppressed by two loops relative to the cutoff, which allows us to push the cutoff to $\sim 10$ TeV while keeping the Higgs light,
\begin{equation}
\delta m_H^2 \sim \left(\frac{\lambda_1^2}{16 \pi^2}\right) \left( \frac{\lambda_2^2}{16\pi^2}\right) \Lambda^2.
\end{equation}
The one-loop quadratic divergences to the Higgs mass-squared from the SM particles are cancelled by new particles which are partners of the SM top quark, gauge bosons and Higgs (Fig.~\ref{fig:cancellation}). Unlike SUSY, these new particles have the same spins as the SM particles, and the relationship of the couplings are dictated by the non-linear realized global symmetry.

\begin{figure}[htbp] 
   \centering
      \includegraphics[width=1.0in]{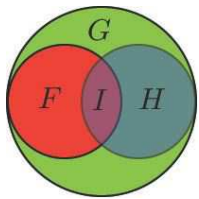} 
   \caption{A global symmetry $G$ is spontaneously broken down to a subgroup $H$. A subgroup $F$ of $G$ is gauged and it is broken down to the intersection of $F$ and $H$, $I = F \cap H$, which is identified as the SM electroweak gauge symmetry. The number of uneaten PNGBs is given by the number of generators of $(N(G)-N(H))-(N(F)-N(I))$. They are identified as the Higgs.}
   \label{fig:coset}
\end{figure}

There are many different little Higgs models based on various global symmetries,  the gauged subgroups, and their breaking patterns. The breaking pattern can be summarized by Fig.~\ref{fig:coset}.

Different types of models can be categorized based on if $G$ and $F$ are simple or product groups. Some representative models are
\begin{itemize}
\item Minimal moose~\cite{Arkani-Hamed:2002qx}: $G/H= SU(3)^2/SU(3), \; F= [SU(2)\times U(1)]^2$.
\item Littlest Higgs~\cite{Arkani-Hamed:2002qy}: $G/H= SU(5)/SO(5),\; F=[SU(2)\times U(1)]^2$.
\item Simple group little Higgs~\cite{Kaplan:2003uc}: $ G/H=[SU(3)/SU(2)]^2$, $F=SU(3)[\times U(1)]$.
\end{itemize}
They all have somewhat different particle contents and spectra, but the generic spectrum for a little Higgs theory looks as follows. At $100- 300$ GeV we have the Standard Model with one or two Higgs doublets. At $\sim 1$ TeV which corresponds to the global symmetry breaking scale, $f$, there are new fermions (partners of the top quark), gauge bosons (partners of the SM gauge bosons), and scalars (partners of the Higgs). The cutoff of the effective theory is at $\Lambda \sim 4\pi f \sim 10$ TeV where additional new physics will come in. 

A major motivation behind the delicate construction of the little Higgs model is the current strong constraints from the electroweak precision data. At energies below the scale of the new physics, the effects of the new physics can be represented by some higher dimensional operators. So far there is no significant deviations from the SM predictions in the electroweak precision measurements, which puts strong bounds on the sizes of these higher dimensional operators. Many of them need to be suppressed by a scale of 10 TeV or higher~\cite{Barbieri:1999tm}. As a result, any new physics at TeV scale has to pass the test of electroweak constraints by not generating large effects in the electroweak observables. The strongest constraints often come from the Peskin-Takeuchi $S$ and $T$ parameters~\cite{Peskin:1990zt,Peskin:1991sw}, 4-fermion interactions, and $Z\to b\bar{b}$ vertex corrections

As the cutoff is pushed up to the 10 TeV scale, the contributions from the cutoff physics to the electroweak observables are mostly safe. However, the new particles at the 1 TeV scale may still induce large corrections to the electroweak observables. The contributions to the $T$ (or $\rho$) parameter can be suppressed by a custodial $SU(2)$ symmetry, hence the models based on $SO(5)$ is favored over  $SU(3)$. On the other hand, the TeV scale particles often gives large contributions to the $S$ parameter and 4-fermion interactions in generic little Higgs models. As a result, electroweak constraints generally require the symmetry breaking scale $f$ to be larger than a few TeV, which would re-introduce the fine-tuning problem~\cite{Hewett:2002px,Csaki:2002qg,Csaki:2003si,Han:2003wu,Gregoire:2003kr,Kilic:2003mq,Marandella:2005wd,Han:2005dz}.

One way to avoid the electroweak precision constraints is to reduce the couplings between the SM fermions and the TeV scale particles. In particular, one can introduce a new symmetry ($T$-parity) into the little Higgs theories~\cite{Cheng:2003ju,Cheng:2004yc,Low:2004xc}. Under the $T$-parity, all SM fields are even and most of the new particles at the 1 TeV scale (those would induce large corrections to electroweak observables) are odd. The $T$-odd particles have to appear at least in pair in any interactions. As a result, there cannot be any contribution to the electroweak observables by exchanging the new particles at the tree level. The leading contributions will be loop suppressed and hence safe from the precision constraints. The $T$-parity can be imposed in many little Higgs models based on symmetric spaces.\footnote{Recently there is a paper discussing the possibility that $T$-parity is broken by the Wess-Zumino term~\cite{Hill:2007zv}. However, this is a UV completion question. It is not difficult to construct models in which the $T$-parity is an exact symmetry in the UV theory, then it will not be violated by the Wess-Zumino term.}

The introduction of $T$-parity has dramatic phenomenological consequences in addition to curing the fine-tuning problem of the little Higgs theories. First, the lightest $T$-odd particle is stable, and can be a good dark matter candidate if it is neutral. Second, the $T$-odd particles need to be paired produced at the colliders and therefore the search strategies are complete different from the case without $T$-parity. After produced, the $T$-odd particles cascade decay  down to the lightest $T$-odd state which escape the detector if it is neutral. Typical collider signals will be jets/leptons + missing energy, similar to SUSY with a conserved $R$-parity.

\subsection{Gauge-Higgs Unification}

One way to protect the squared mass of a scalar field from quadratic divergence is to identify the scalar as the extra components of some gauge field in extra dimensions. Then the scalar mass is protected by the gauge symmetry in extra dimensions, and only receives a finite contribution when the extra dimensions are compactified. This idea has also been around for a long time but only receives a lot of attention and extensive studies recently~\cite{Dvali:2001qr,Hall:2001zb,Antoniadis:2001cv,Csaki:2002ur,Burdman:2002se,Haba:2002vc,Gogoladze:2003bb,Scrucca:2003ut,Haba:2004qf}. To get the quantum number of the Higgs doublet, one can consider either an $SU(3)$ bulk gauge group broken down to $SU(2)(\times U(1))$ by boundary conditions, or $SO(5)$ broken down to $SO(4)$. The Higgs is associated with the broken generators. Phenomenologically the latter is favored because it contains the custodial $SU(2)$ symmetry which protects the $\rho$ parameter.

The extra dimensions can be either flat or warped. In the case of warped extra dimensions, the model has a dual description that Higgs arises as the PNGB of a spontaneously broken global symmetry of some strongly coupled conformal field theory (CFT)~\cite{Contino:2003ve,Agashe:2004rs}. 

Although gauge-Higgs unification seems to involve very different theoretical constructions from the little Higgs models. They share some similar features. The Higgs mass is protected by a shift symmetry and they are often based on the same group structure. In fact, their low energy phenomenologies can be quite similar too and there can be a unified approach to study them, which will be discussed in the next subsection.

\subsection{Little M-theory}

As we discussed in the previous sections, there are many different little Higgs models as well as models based on extra dimensions such as the gauge-Higgs unification models.
The vast amount of possibilities introduces a practical problem: which models deserve more detailed studies and can they well represent the class of models to which they belong? 

In fact, many of these new models have similar new particles (new quarks, new spin-1 bosons). It is desirable if there is a unified approach for the LHC studies, and in particular, if there are models that can interpolate among various models and cover most of the features of the non-SUSY theories. Indeed, this is possible because most of the non-SUSY models can be well represented or approximated by some moose diagrams at low energies. For example, a gauge field propagating in extra dimensions can be deconstructed into a series of four-dimensional gauge groups broken down to the diagonal group~\cite{Arkani-Hamed:2001ca,Hill:2000mu}. This works for both flat and curved extra dimensions as the warp factor can easily be represented by different Goldstone decay constants of the link fields~\cite{Cheng:2001nh,Sfetsos:2001qb,Randall:2002qr}.

\begin{figure}[htbp] 
   \centering
  \includegraphics[width=3.0in]{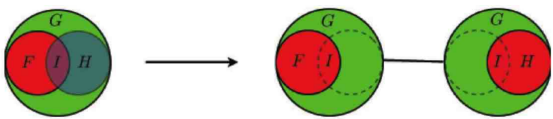} 
   \caption{The graphical representation of converting a non-moose theory into a moose theory}
   \label{fig:ccwz}
\end{figure}

For little Higgs theories, some models are also based on moose diagrams,  e.g., minimal moose model. On the other hand, there are also models containing simple global or gauged groups such as the littlest Higgs and simple group little Higgs models. At the first sight, they are not described by moose diagrams. However, by using the construction of Callan, Coleman, Wess, and Zumino (CCWZ)~\cite{Coleman:1969sm,Callan:1969sn} and related ideas such as hidden local symmetry~\cite{Bando:1984ej,Bando:1985rf,Bando:1987ym,Bando:1987br}, and AdS/CFT correspondence plus deconstruction, they can all be converted into moose models. The procedure is to extend the global group and its symmetry breaking pattern to $(G\times G)/G$, with a subgroup $F$ gauged in the first $G$ group and a subgroup $H$ gauged in the second $G$ group~\cite{Thaler:2005kr}. The gauged subgroups are broken down to the SM electroweak group $I$. This is described in Fig.~\ref{fig:ccwz}. One can easily show that the number of uneaten PNGB's are the same as before. The extra degrees of freedom added in this extension decouple if one takes the gauge coupling of the $H$ subgroup to be large. In this limit, one recovers exactly the same low energy effective theory as the original model, but now the new model is described by a moose diagram.

\begin{figure}[htbp] 
   \centering
   \includegraphics[width=3.0in]{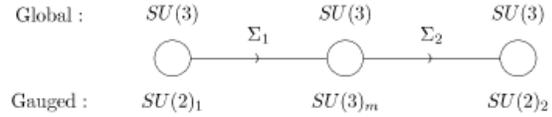} 
   \caption{The three-site moose modell}
   \label{fig:simple}
\end{figure}

One can easily apply this procedure to various little Higgs models. After the conversion, one finds that many different models can often be represented by the same moose diagram in different limits. For example, consider the three-site moose diagram with an $SU(3)$ global symmetry on each site. The whole $SU(3)$ is gauged at the middle site while at the two boundary sites only $SU(2)[\times U(1)]$ subgroups are gauged (Fig.~\ref{fig:simple}).
The three-site moose model is quite versatile and can describe several different-looking models by taking various limits. The simple group little Higgs limit is obtained by taking the gauge couplings of the two $SU(2)$'s on the boundaries to infinity. On the other hand, if we keep the gauge couplings of the two $SU(2)$'s finite but take the gauge coupling of the middle $SU(3)$ to infinity, we can integrate out the middle site and reduce it to the minimal moose little Higgs model. The $T$-parity can easily be implemented by taking the gauge couplings of the two $SU(2)$'s and the two decay constants associated with the links equal. Furthermore, it can also be a good low energy description of the holographic PNGB Higgs model~\cite{Contino:2003ve,Agashe:2004rs}. The holographic PNGB Higgs model is based on the AdS/CFT correspondence. In the AdS description, there is an $SU(3)$ gauge symmetry in the bulk while only an $SU(2)$ is gauged on each of the UV and IR branes. By a simple deconstruction, we can see that the three-site moose gives a good approximation of the first few KK modes of this model. The model based on $SU(3)$ does not have a custodial $SU(2)$ symmetry to protect the $\rho$ parameter, but it can easily be fixed by replacing $SU(3)$ with $SO(5)$.

Of course, different models have very different UV behaviors. The three-site model or any other constructions based on the idea discussed in this subsection can only be the low energy approximations of these various models. However, this construction can be a useful tool for the phenomenological studies in the LHC era~\cite{Cheng:2006ht}. The reason is that the precision electroweak constraints indicate that the scale of the strong dynamics may be out of the reach of the LHC, and in the case of extra dimensions only a few KK modes may be discovered at the LHC at most, so we will only be able to test the low energy part of a fundamental theory. At low energies (accessible to the LHC), most non-SUSY models can be well represented by some simple moose models, as moose diagrams are just a convenient way to categorize spin-1 and spin-0 degrees of freedom. As we found, many models can even be described  as various limits of the same moose diagram. Furthermore, this construction also reveals a much larger model space for new physics as it interpolates among various models proposed before. There is a continuous distribution of viable models simply by changing the parameters of the ``unified'' model.  


\subsection{Low-energy Effective Lagrangian for a Strongly Interacting Light Higgs}

If the scale of the new particles or strong dynamics responsible for the composite light Higgs is somewhat higher than 1 TeV, they may not be immediately accessible and the only particle we see is the light Higgs. An important question is how we can test this scenario and distinguish it from the Standard Model. A low-energy effective Lagrangian approach for such a strongly-interacting light Higgs (SILH) has been investigated by Giudice et al~\cite{Giudice:2007fh}. Some more specific results can be obtained beyond the general higher-dimensional operator analysis for this class of models.

The strongly coupled sector can be characterized by two parameter, coupling $g_{\rho} (> g_{SM})$ and the mass scale of the new states (other than the light Higgs) from the strong dynamics $m_{\rho}$. The symmetry breaking scale $f$ is related to them by $m_{\rho} = g_{\rho} f$. Integrating out the strong sector, the low-energy effective Lagrangian can be expressed in terms of the expansions of $H/f$ and $\partial/ m_{\rho}$. The resulting higher-dimensional operators can be divided into the following two categories:
\begin{itemize}
\item Genuine strong operators, which is sensitive to the symmetry breaking scale $f$;

\item Form factor operators, which is sensitive to the resonance scale $m_{\rho}$.
\end{itemize}

The genuine strong operators include
\begin{eqnarray}
&& \frac{c_H}{2f^2} \partial^\mu \left(H^\dagger H\right) \partial_\mu \left( H^\dagger H\right), \nonumber \\
&& \frac{c_T}{2f^2} \left( H^\dagger \stackrel{\leftrightarrow}{D^\mu} H\right)\left( H^\dagger \stackrel{\leftrightarrow}{D_\mu} H\right), \nonumber \\
&& \frac{c_y y_f}{f^2} H^\dagger H \bar{f}_L H f_R ,\nonumber \\
&& \frac{c_6 \lambda}{f^2} \left(H^\dagger H\right)^3 .
\end{eqnarray}
The first two terms are fixed by the $\sigma$-model structure and the last two terms come from both the $\sigma$-model structure and the resonances at $m_{\rho}$. The second term contributes to the $T$-parameter and represents the custodial $SU(2)$ breaking effect. It is strongly constrained experimentally and can be suppressed by the presence of a custodial $SU(2)$ symmetry of the strong sector. It can hence be ignored for most reasonable models. The first term rescales the Higgs kinetic term after substituting the Higgs by its vev. It modifies the Higgs coupling to the SM fields. As a result, unitarity is not exactly restored by the Higgs along but also the heavy resonances in the strong sector. This provides a model-independent test of the compositeness nature of the Higgs field.

Form factor operators involving the Higgs and the gauge fields at dimension-6 are
\begin{eqnarray}
&& \frac{i c_W g}{2 m_{\rho}^2} \left( H^\dagger \sigma^i \stackrel{\leftrightarrow}{D^\mu} H\right) (D^\nu W_{\mu\nu})^i \nonumber \\
&+& \frac{ic_B g'}{2 m_\rho^2}  \left( H^\dagger \stackrel{\leftrightarrow}{D^\mu} H\right) (\partial^\nu B_{\mu\nu}) \nonumber \\
&+&  \frac{i c_{HW} g}{16 \pi^2 f^2} ( D^\mu H)^\dagger \sigma^i ( D^\nu H) W^i_{\mu\nu} \nonumber \\
&+& \frac{i c_{HB} g'}{16 \pi^2 f^2} (D^\mu H)^\dagger (D^\nu H) B_{\mu\nu} \nonumber \\
&+& \frac{c_\gamma g'^2}{16\pi^2 f^2} \frac{g^2}{g_\rho^2} H^\dagger H B_{\mu\nu} B^{\mu\nu} \nonumber \\
&+& \frac{c_g g_S^2}{16\pi^2 f^2} \frac{y_t^2}{g_\rho^2} H^\dagger H G^a_{\mu\nu} G^{a\mu\nu}.
\end{eqnarray}
General analysis shows that the $c_H$ and $c_y$ terms are the most important ones for the LHC studies of this class of models.

\section{Heavy Composite Higgs and No Higgs Scenarios}

\subsection{Heavy Composite Higgs}

In this scenario, Higgs is a composite field produced by some strong dynamics, but there is no symmetry to protect the Higgs mass. Therefore, the Higgs mass is expected to be heavy, close to the scale of strong dynamics, unless some fine-tuning is involved to make it light. Some old examples are the top condensate model and its variations~\cite{Nambu:1988,Miransky:1989ds,Miransky:1988xi,Bardeen:1989ds,Hill:1991at}.

\begin{figure}[htbp] 
   \centering
   \includegraphics[width=2.5in]{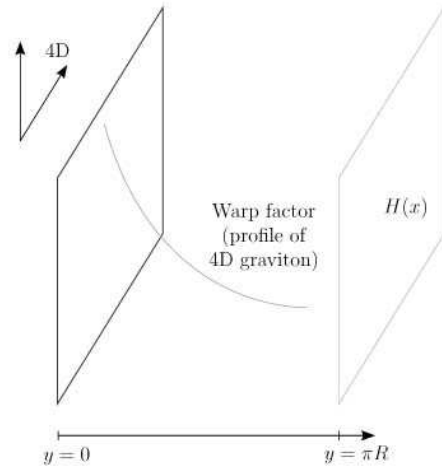} 
   \caption{The Randall-Sundrum model with an warped extra dimension. }
   \label{fig:RS1}
\end{figure}

Recent developments in extra-dimensional theories and AdS/CFT correspondence provide a new tool to study this kind of models. In particular, the Randall-Sundrum model (RS1)~\cite{Randall:1999ee} gives a calculable extra-dimensional dual description of such a scenario with a strongly coupled CFT. It is based on a warped extra dimension cut off by an UV (Planck) brane and an IR (TeV) brane (Fig.~\ref{fig:RS1}). The Higgs field localized at (or near) the IR brane is interpreted as a bound state of the strongly coupled CFT in the four-dimensional picture.

To satisfy the electroweak precision constraints and to address fermion mass hierarchies, it is favorable to have gauge fields propagating in the bulk too~\cite{Agashe:2003zs}. To protect the $\rho$ parameter the CFT should contain the custodial symmetry, which implies that we need to have the  $SU(2)_L \times SU(2)_R$ gauge group in the bulk. SM fermion masses may be explained by the localizations of the fermions. Light generations are localized towards the UV brane so that they have small overlaps with the Higgs. In the 4-dimensional picture this means that they are (mostly) fundamental fields. The top quark should be localized near the IR brane to accommodate the large top Yukawa coupling. In the 4-dimensional picture this means that top quark is a (partially) composite state from the strongly coupled CFT.

\subsection{No Higgs Scenario}

If electroweak symmetry is broken by strong dynamics, it is also possible that there is no Higgs particle and the longitudinal $WW$ scattering is unitarized by some other states. Technicolor theories are the original models without the Higgs~\cite{Weinberg:1979bn,Susskind:1978ms}. The longitudinal $WW$ scattering is unitarized by some resonances such as techni-rhos in these models. Because of the strong dynamics, it is difficult to make precise predictions in these models, though naive estimates extrapolating from QCD make these models disfavored by the electroweak precision constraints.

\begin{figure}[htbp] 
   \centering
   \includegraphics[width=3.0in]{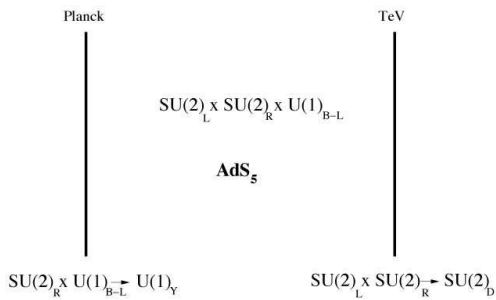} 
   \caption{The 5D Higgsless model in warped space }
   \label{fig:higgsless}
\end{figure}

Similar to the composite Higgs scenario, the warped extra dimensions and AdS/CFT correspondence allows an alternative and calculable description of certain Higgsless models~\cite{Csaki:2003dt,Csaki:2003zu}. The setup is similar to RS1 with gauge field in the bulk, except that there is no Higgs field. Electroweak symmetry is broken by the combination of the boundary conditions at UV and IR branes (Fig.~\ref{fig:higgsless}). In this picture, the longitudinal $WW$ scattering is unitarized  by the KK gauge bosons of the SM gauge fields.

This model can also be viewed as a limit of taking the vev of the Higgs localized on the IR brane to infinity in the RS1 scenario. From this point of view, there can also be models which interpolating between the heavy Higgs and Higgsless scenarios. There will be a Higgs boson in these interpolating models, but its couplings to $W$ and $Z$ gauge bosons are suppressed~ \cite{Cacciapaglia:2006mz}. The unitarization of the longitudinal $WW$ scattering is shared by the Higgs boson and the KK gauge bosons.

\subsection{Electroweak Constraints}

As we discussed before, electroweak precision data provide strong constraints on any new physics at the TeV scale, including the heavy composite Higgs and Higgsless models. The strongest constraints come from the $S$, $T$ parameters, four-fermion interactions and $Z\to b \bar{b}$ vertex corrections. Because we have the calculable 5D pictures for these models, their corrections to the electroweak observables can be calculated reliably. Contribution to the $T$ parameter can be suppressed by including a custodial symmetry (i.e., gauging $SU(2)_L\times SU(2)_R$ in the 5D dual description). Contribution to the $S$ parameter is positive if the SM fermions are localized on the UV brane, in agreement with the estimates in Technicolor theories. To satisfy the constraint on the $S$ parameter, there are a couple possibilities.
\begin{itemize}
\item  If there is a Higgs boson, one can raise the KK gauge boson masses (the compositeness scale) at the expense of more fine-tuning.

\item In Higgsless models, the KK gauge bosons have to be around 1 TeV because they are responsible for 
unitarizing the $W_L W_L$ scattering. One can reduce their couplings to the light generations of SM fermions by choosing a near-flat profile in the bulk for the light fermions.
\end{itemize}

These solutions also alleviate the constraints on 4-fermion interactions. However, to have large enough top quark mass, top quark needs to be near the IR brane. In the traditional embedding of the left-handed top-bottom doublet, $(t_L, \, b_L) \sim (\bf{2},\, \bf{1})$ under $SU(2)_L\times SU(2)_R$, $(t_L,\, b_L)$ mixes with KK states transforming as $(\bf{1},\, \bf{2})$, which induces large corrections to $Z\to b\bar{b}$. This problem can be solved by choosing a different embedding, $(t_L, \, b_L) \sim (\bf{2},\, \bf{2})$ with a custodial symmetry $SU(2)_L\times SU(2)_R \times P_{LR}$~\cite{Agashe:2006at}.

\subsection{Deconstruction}

Similar to the light Higgs scenario, at energy scales accessible to the LHC, the heavy Higgs and Higgsless models can be approximated by deconstruction with only a few sites. Such deconstructions are useful for LHC phenomenology studies as the prototype models for a generic class of models. They have been constructed for both heavy composite Higgs models~\cite{Contino:2006nn} and Higgsless models~\cite{SekharChivukula:2006cg}.

\section{Conclusions}

Recent new ideas such as extra dimensions, AdS/CFT correspondence, and collective symmetry breaking have provided us many new tools for model building. As a result, many new models of electroweak symmetry breaking have been constructed, including little Higgs, heavy composite Higgs, and Higgsless models. Although different names suggest different ideas and emphases, the model space is actually continuous and there are models which interpolating among various scenarios. The new theoretical developments also allow us to have a new and uniform way to understand and implement many old ideas scattered at various places.

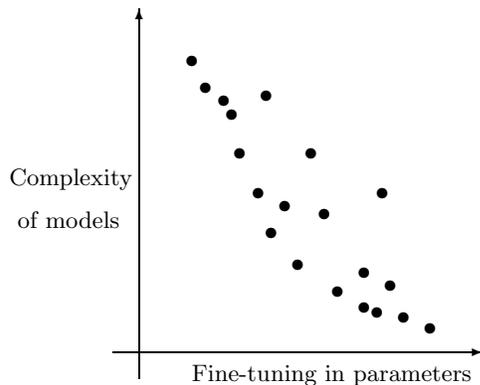
\begin{figure}[htbp] 
   \centering
\begin{picture}(200,150)(0,0)
\put(40,15){\vector(1,0){140}}
\put(50,5){\vector(0,1){140}}
\put(70,4){Fine-tuning in parameters}
\put(1,78){Complexity}
\put(1,62){ of models}
\put(160,24){\circle*{4}}
\put(150,28){\circle*{4}}
\put(145,40){\circle*{4}}
\put(142,75){\circle*{4}}
\put(140,30){\circle*{4}}
\put(135,45){\circle*{4}}
\put(135,32){\circle*{4}}
\put(125,38){\circle*{4}}
\put(120,67){\circle*{4}}
\put(115,90){\circle*{4}}
\put(110,48){\circle*{4}}
\put(105,70){\circle*{4}}
\put(100,60){\circle*{4}}
\put(95,75){\circle*{4}}
\put(88,90){\circle*{4}}
\put(98,112){\circle*{4}}
\put(85,105){\circle*{4}}
\put(82,110){\circle*{4}}
\put(75,115){\circle*{4}}
\put(70,125){\circle*{4}}

\end{picture}
   \caption{Tuning vs complexity in beyond SM model space}
   \label{fig:tuning}
\end{figure}

Among the large number of new models alternative to SUSY appeared recently, it is hard to argue that any particular model stands out as the most probable one for new physics at the TeV scale. A major reason is due to the electroweak precision data. To satisfy the strong electroweak constraints, one often needs to trade between fine-tuning and excessive model-building. The models often end up quite complicated if one tries to avoid any tuning. The situation can be schematically represented in Fig.~\ref{fig:tuning}. The horizontal axis represents the amount of tuning in model parameters required and the vertical axis represents the complexity of the models. There is no simple model without any tuning remaining in the valid model space.

Given so many as good (or bad) models, it is highly desirable to have some unified approaches to them. This is made possible by effective Lagrangians and moose diagrams. Most of these new models predicts new vector particles (KK gauge bosons or techni-rhos) and new fermions associated with the third generation SM fermions. They can be easily represented by some moose diagrams. If such new states are discovered at the LHC, constructing a moose model can be a first step towards uncovering the underlying theory. 

%
%

\end{document}